\documentclass[runningheads]{llncs}
\usepackage{graphicx}
\usepackage{orcidlink}
\usepackage{todonotes}
\usepackage{longtable}
\usepackage{hyperref}

\usepackage{xcolor}
\usepackage{colortbl}
\usepackage{comment}
\usepackage{adjustbox}
\usepackage[T1]{fontenc}
\usepackage{acronym}
\usepackage{xspace}
\newcommand{\SymbPC}{\textsc{SymboleoPC}\xspace}
\newcommand{\SymbSC}{\textsc{Symboleo2SC}\xspace}
\newcommand{\SymbNLP}{\textsc{SymboleoNLP}\xspace}
\newcommand{\SymbWeb}{\textsc{SymboleoWeb}\xspace}

\usepackage{array}
\usepackage{rotating}
\usepackage{multirow}
\usepackage{graphicx}
\usepackage{makecell}
\usepackage[utf8]{inputenc}
\usepackage[final]{listings}
\newcommand{\Symbo}{\textsc{Symboleo}\xspace}
\renewcommand\normalsize{\fontsize{10pt}{12pt}\selectfont}

\lstdefinelanguage{Symboleo}{
language = C++,
morekeywords = { 
Domain, endDomain, Contract, Declarations, Preconditions, Postconditions, Obligations, Surviving Obligations, Powers, Power, Obligation, Number, Constraints, endContract, isA, isAn, Enumeration, with, String, Date, Boolean, Asset, Event, Role, Env, or, and, not, true, false, O, P, Suspended, Resumed, Discharged, Terminated, Happens, HappensBefore, HappensAfter, HappensWithin, Occurs, IsEqual, IsOwner, CannotBeAssigned, Triggered, Activated, Exerted, Expired, Fulfilled, Discharged, Violated, FulfilledObligations, RevokedParty, AssignedParty, Rescinded, Create, UnsuccessfulTermination, Active, InEffect, Suspension, SuccessfulTermination, Discharge, Violation, Fulfillment, UnsuccessfulTermination, String.concat, Math.pow, Math.abs, WhappensBefore, ShappensBefore UnAssign, Form, Rescission
},
morecomment=[l]{--},
backgroundcolor=\color{lightgray!20},
keywordstyle=\color{blue!90}\bfseries,
basicstyle=\scriptsize\ttfamily,
commentstyle=\color{green!60!black},
numbers=left,
numberstyle=\tiny\ttfamily,
numbersep=2.5em,
showstringspaces=false,
breaklines=true,
frame=lines,
framexleftmargin=0.5em,
xleftmargin=0em,
tabsize=2,
}

\definecolor{lightgreen}{rgb}{0.88, 1.0, 0.88} 
\begin{document}
\title{A Web-Based Environment for the Specification and Generation of Smart Legal Contracts \thanks{Partially funded by NSERC's Strategic Partnership Grant \textit{Middleware Framework and Programming Infrastructure for IoT Services}, by SSHRC's Partnership Grant \textit{Autonomy Through Cyberjustice Technologies}, and by the ORF-RE project \textit{CyPreSS: Software Techniques for the Engineering of Cyber-Physical Systems}.}}

\titlerunning{A Web-Based Environment for Smart Legal Contracts}

\author{Regan Meloche\inst{1}\orcidlink{0009-0004-2418-1990} \and
Durga Sivakumar\inst{1}\orcidlink{0009-0001-0469-3532} \and
Amal A. Anda\inst{1}\orcidlink{0000-0001-7851-0199} \and
Sofana Alfuhaid\inst{1,2}\orcidlink{0009-0006-9087-3041} \and
Daniel Amyot\inst{1}\orcidlink{0000-0003-2414-1791} \and
Luigi Logrippo\inst{1,3}\orcidlink{0000-0001-8804-0450} \and
John Mylopoulos\inst{1}\orcidlink{0000-0002-8698-3292}}
\authorrunning{R. Meloche et al.}
%
\institute{University of Ottawa, Ottawa, ON, K1N 6N5, Canada\\ 
\email{\{rmeloch2, dsiva066, aanda, salfu014, damyot, jmylopou\}@uottawa.ca}\\
\url{https://sites.google.com/uottawa.ca/csmlab/} 
\and
King Abdulaziz University, Jeddah, Saudi Arabia\\
\and
Universit{\'e} du Qu{\'e}bec en Outaouais, Gatineau, QC, J8X 3X7, Canada\\
\email{luigi@uqo.ca}\\
}
\maketitle              
\begin{abstract}
Monitoring the compliance of contract performance against legal obligations is important in order to detect violations, ideally, as soon as they occur. Such monitoring can nowadays be achieved through the use of smart contracts, which provide protection against tampering as well as some level of automation in handling violations. However, there exists a large gap between natural language contracts and smart contract implementations. This paper introduces a Web-based environment that partly fills that gap by supporting the user-assisted refinement of \Symbo specifications corresponding to legal contract templates, followed by the automated generation of monitoring smart contracts deployable on the Hyperledger Fabric platform. This environment, illustrated using a sample contract from the transactive energy domain, shows much potential in accelerating the development of smart contracts in a legal compliance context. 

\end{abstract}
\section{Introduction}
Legal contracts describe obligations and powers that apply to business transactions between parties playing roles such buyers and sellers, or service providers and consumers. Most often, they are redacted in Natural Language (NL). Violations of obligations may enable defined powers such as premature contract termination or penalties. They can also lead to costly conflict resolution mechanisms, including litigation in courts. Monitoring contract performance is important to detect such violations as early as possible, in order to favor potential alternative courses of actions and avoid litigation. 

Nowadays, legal contracts can benefit from various automation strategies and technologies. For example, smart contracts are programs intended to partially automate legal contracts and monitor them for compliance with relevant terms and conditions. However, this paper uses a generic definition of smart contracts based on Szabo's~\cite{Szabo1997}, which is independent of the blockchain-based technologies commonly referenced in the past 15 years. 

Contract templates capture standard clauses that enable their reuse (e.g., by selecting clauses), customization (e.g., through parameters), and refinement (e.g., by modifying clauses) for particular contexts. Such templates are particularly useful in situations where the type of business transaction targeted is well understood. One issue here is that refinement of a template may lead to an invalid contract (e.g., due to newly introduced inconsistencies) or to the unexpected dissatisfaction of desirable properties that the original template satisfied. Another issue is that the monitoring mechanisms supporting the original template may need to be updated to reflect such refinements.

One solution avenue for both problems is the use of an intermediate formalization of legal templates that supports both property verification and the generation of smart contracts that monitor legal contract executions~\cite{Soavi2022}. For example, the \Symbo language~\cite{symboleo} can be used to specify contract templates in terms of parameterized obligations and powers, while enabling formal verification of properties~\cite{Parvizimosaed2022-SymboleoPC} and the automated generation of monitoring smart contracts~\cite{rasti2024-Symboleo2SC}. Although \Symbo is supported by an Eclipse-based editor and automated code generators~\cite{symboleo}, the creation of formal specifications for contract templates and their refinements are currently done \textit{manually}. Also, the Eclipse Integrated Development Environment (IDE)~\cite{Eclipse} is a desktop environment that requires many components to be installed locally, which is a major barrier to entry for many potential \Symbo users. 

In an effort to simplify and partially automate the generation and modification of \Symbo specifications, this paper provides two main contributions:
\begin{itemize}
    \item A Web-based tool (\SymbNLP, \url{https://bit.ly/SymboleoNLP}) that handles part of the formalization of NL templates to \Symbo specifications, with a focus on temporal and conditional refinements.
    \item A Web-based IDE (\SymbWeb, \url{https://bit.ly/Symboleo_Web}) that takes the generated \Symbo, enables intelligent editing and validation, and converts it to executable smart contracts in JavaScript for the Hyperledger Fabric distributed ledger platform (\url{https://www.hyperledger.org/projects/fabric}). \SymbWeb simplifies the access to \Symbo tools.
\end{itemize}

This composite Web-based environment is illustrated here using a \textit{transactive energy} (TE) contract template example. TE markets already use smart contracts and are eager to support small-scale but automated contracts for trading energy between prosumers (i.e., consumers who also produce and sell energy). 

The rest of this paper is as follows. Section~\ref{RelatedWork} first presents related work in that area. Section~\ref{FromNLtoSymboleo} describes how \SymbNLP helps generate \Symbo specifications from refined NL templates, while Section~\ref{FromSymbToSC} illustrates how \SymbWeb supports editing and code generation. An analysis of different refinements to a TE contract template in terms of their impact on the size of the \Symbo specifications and JavaScript code generated is provided in Section~\ref{Analysis}. Finally, Sections~\ref{Discussion}
and~\ref{Conclusion} present a discussion and our conclusions, respectively.

\section{Related Work}
\label{RelatedWork}
There exist several formal specification languages for legal contracts~\cite{Soavi2022} and for smart contracts~\cite{Tolmach2022}. Some support a certain degree of code generation from abstract specifications, while some support the formal verification of properties~\cite{Krichen2022,Tolmach2022}. \Symbo is unique since it enables both i)~formal verification of desirable properties expressed in Linear Temporal Logic or in Computation Tree Logic~\cite{Parvizimosaed2022-SymboleoPC}, as well as ii)~the automated generation of smart contracts (using the \SymbSC~\cite{rasti2024-Symboleo2SC} tool) deployable on the Hyperledger Fabric platform. 

\Symbo exploits a legal contract ontology with concepts such as obligation, power, party, role, asset, and event~\cite{symboleo}. The language enables modelers to specify contracts by defining a domain model that extends this ontology, and then the legal obligations and powers that express legal clauses and terms.

\Symbo has been used to specify different types of supply chain contracts that can benefit from automated monitoring of performance. This includes the transactive energy domain~\cite{TESC-2020}, where energy is the asset being sold from prosumers to buyers. TE contracts have also been coded manually by others as smart contracts running over different blockchain platforms~\cite{Khan2023-SC-Energy,Kirli2022-SC-Energy}.

One key issue is the generation of formal specifications from NL contracts. For the \Symbo ontology, some preliminary work was done by Soavi et al.~\cite{SoaviContrattoA-2022}, but with limited automation. Meloche et al.~\cite{regan-thesis,Meloche2023RE} recently provided a higher level of automation in the generation of \Symbo specifications, but only for certain categories of legal contracts \textit{templates}. An unrefined template (\textit{T}) is converted manually to \Symbo (specification \textit{S(T)}), and then manual refinements (\textit{C}) of the original template, expressed using a controlled natural language (CNL) supported by a dynamic grammar, are automatically transformed into valid refinements of the specification (\textit{S(C)}). We extend this work with a more complete translation and a web-based interface implemented in the \SymbNLP tool.

Tateishi et al.~\cite{Tateishi2019} proposed a technique and a toolset for automatically generating a smart contract from a contract written in a CNL using a template-based approach. For a given contract domain, a contract template is manually created, which contains parameters that can be customized using the CNL. Given the template and the customizations, a formal model is automatically generated in a domain-specific language. The formal model can then be translated into smart contract code. The approach and goals are similar to \SymbNLP, but a key difference is the scope of customized refinements. In this related work, the allowed customizations are specified individually for each parameter, meaning that the preparation can be quite tedious. \SymbNLP's approach is to have a more general CNL that can be used with a large variety of template parameters. This important difference makes the initial set up process much easier.

Dwivedi et al.~\cite{Dwivedi2021} developed a smart contract language (SCL) for decentralized autonomous organizations. SCL shares some ontological and technical similarities with \Symbo. For example, both approaches address the ability for certain norms to trigger the creation of other norms (e.g., the failure to complete an obligation can trigger the instantiation of a power to terminate the contract). A markup language to specify contracts (SCLML) is based on this ontology and can be converted to Solidity code for a smart contract implementation. 

\section{From Contract Templates to \Symbo}
\label{FromNLtoSymboleo}

We can illustrate the process of formalizing refinements on a NL contract into \Symbo using a simple Transactive Energy (TE) contract, shown in Table~\ref{tab:contract}.

\definecolor{myblue}{rgb}{0.95, 0.95, 1}
\rowcolors{1}{myblue}{myblue}
\footnotesize
\begin{longtable}{ p{\linewidth}}
\hline
This agreement is entered into effect as of <date>, between <buyer> as Buyer and <prosumer> as Prosumer.
\begin{enumerate}
    \item Prosumer shall dispatch <energy\_qnt> kW of power to the Buyer \textbf{[P1]}.
    \item The power needs to be dispatched to <location>.  
    \item The Buyer shall pay <amount> in CAD to the Prosumer \textbf{[P2]}.
    \item In case of late payment, the Buyer shall pay a late fee equal to <percentage> of the amount owed, and the Prosumer may suspend performance of all of its obligations under the agreement until payment of amounts owed has been received in full.
    \item Any delay in the dispatching of the power, or any problem with the voltage (which must be between <min> V and <max> V), will entitle the Buyer to terminate the Contract.
\end{enumerate}\\
\hline
\rowcolor{white}\caption{\normalsize Simple TE contract with two refinement locations (P1, P2).}
\label{tab:contract}
\end{longtable}
\rowcolors{1}{white}{white}
\normalsize
From this contract, we will consider the first sentence, which specifies an obligation of the prosumer to dispatch an amount of power to the buyer. The obligation specifies that the power delivery must be made within a certain timeframe. The contract template allows for flexibility in choosing the start and end times of this timeframe. However, there may be cases where more flexibility is needed in the contract template. For example, we may want to specify that the power must be delivered \textit{before} a certain date, \textit{after} a certain date, or perhaps a more complex temporal specification (e.g., within 2 weeks of buyer paying the prosumer). \SymbNLP is built to handle these more complex temporal refinements, as well as certain types of conditional refinements. 

The process begins with a broadly-specified NL contract \textit{T} and its corresponding \Symbo specification \textit{S(T)}, which, using our example sentence, are presented as follows:
\vspace{-2.5mm}
\begin{itemize}
    \item \textit{T}: Prosumer shall dispatch <energy\_qnt> kW of power to the buyer [\textbf{P1}]
    \item \textit{S(T)}: {\lstinline[language=Symboleo, basicstyle=\ttfamily\small]|Obligation(Prosumer, Buyer, true, Happens(evt_dispatch_energy))|}
\end{itemize}

Note that in practice, \textit{T} and \textit{S(T)} will correspond to full contracts. We simplify it here to correspond to only one sentence for illustrative purposes. The full \Symbo contract will also include more detailed definitions of the Prosumer and Buyer roles, as well as of \texttt{\small evt\_dispatch\_energy}, which specifies the event that must happen to fulfill the obligation. We omit details for conciseness.

The artifacts \textit{T} and \textit{S(T)} are fed to the \SymbNLP tool, and the contract author can make temporal and conditional refinements on the parameter [P1] in \textit{T}. The refinements that the contract author is allowed to specify come from a carefully-constructed CNL that is based on a dataset of legal contract refinements~\cite{Meloche2023RE,regan-thesis}. These refinements often come in the form of a linguistic \textit{adjunct}, a component of a sentence that adds information, but when removed from the sentence, does not alter the basic sentence structure. Adjuncts often add information about time, space, or method, and they can take various grammatical forms, including prepositional phrases and adverbs. Some examples of adjuncts that may be allowed in the CNL for this refinement include: i)~\textit{between [START\_DATE] and [END\_DATE]}; ii)~\textit{before March 31, 2023}; and iii)~\textit{within 2 weeks of the buyer paying the prosumer}. Note how these two prepositional phrases add extra information to the sentence, but are \textit{not} required for the sentence to be complete; They simply restrict the meaning, which is the function of adjuncts.

\begin{figure}[h!]
\centering
\fbox{\includegraphics[width=\textwidth]{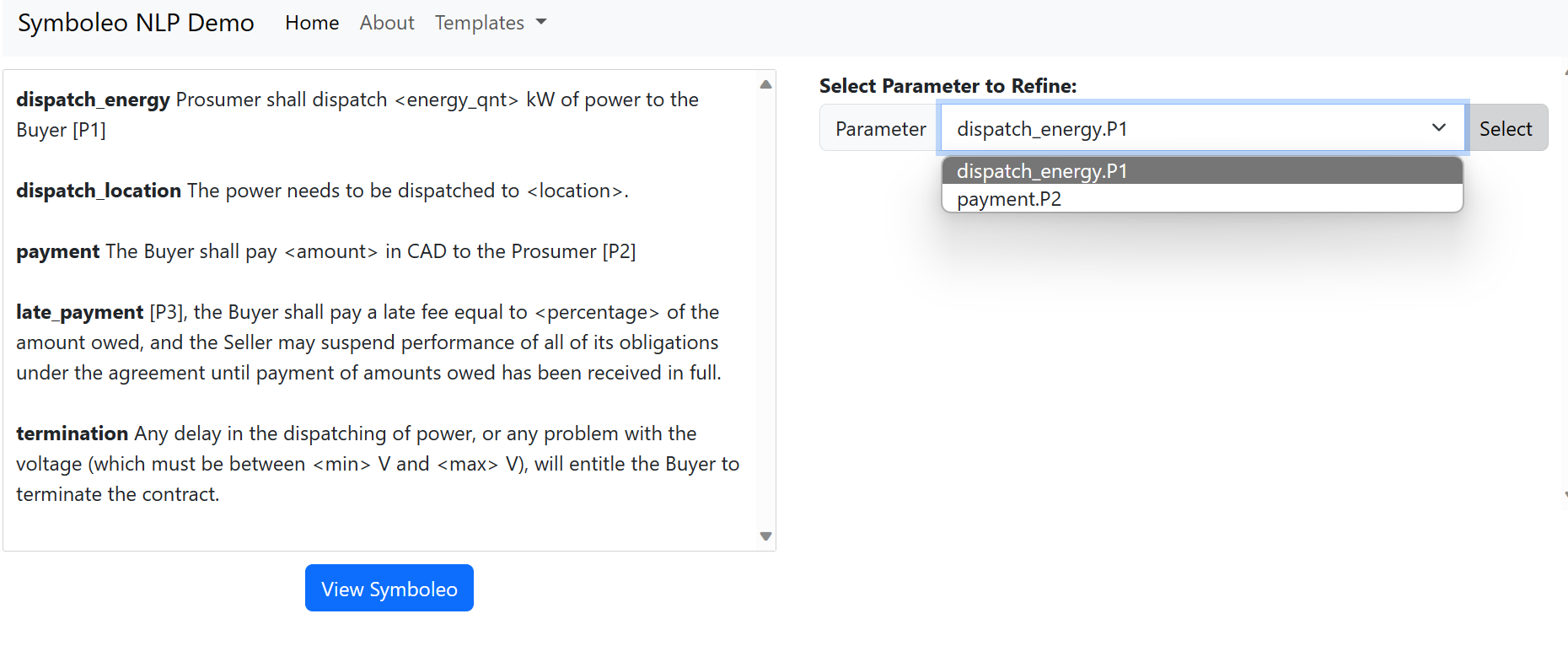}}
\caption{The user selects the parameter to refine.}
\label{fig:s3}
\centering
\end{figure}

\begin{figure}[h!]
\centering
\fbox{\includegraphics[width=0.47\textwidth]{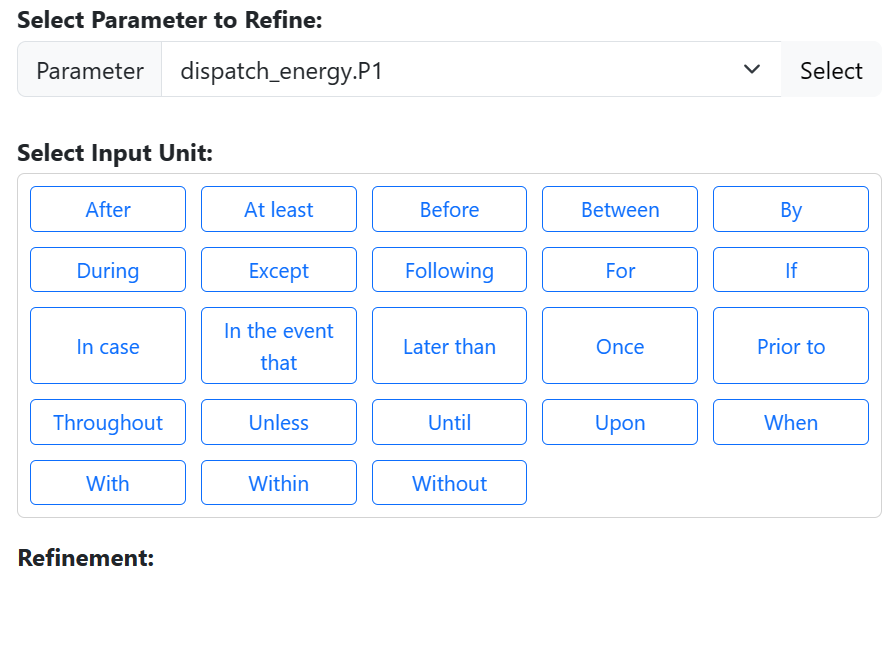}}
\fbox{\includegraphics[width=0.47\textwidth]{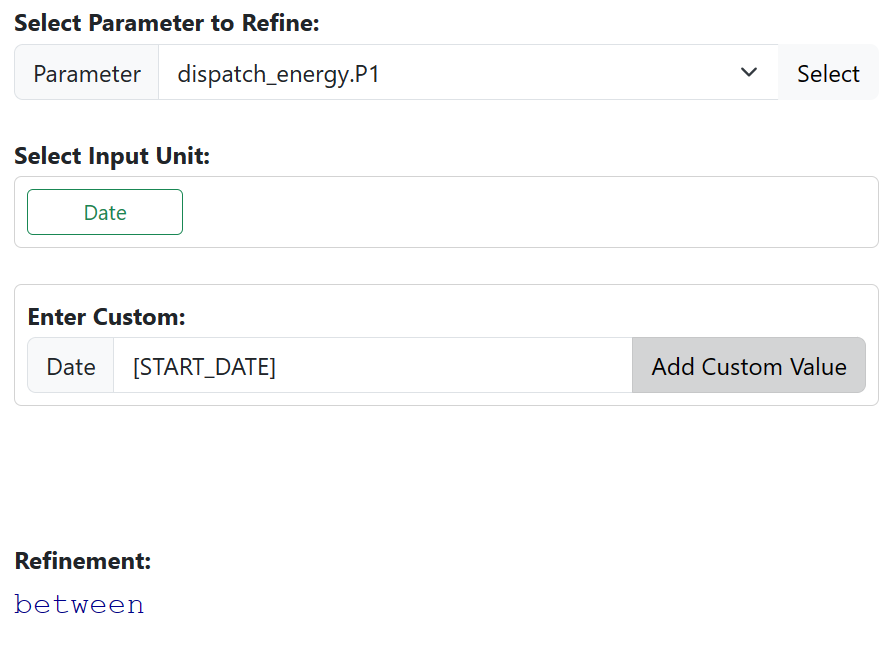}}
\caption{Left: The user enters the refinement according to the text values allowed by the CNL. In some cases, the user is given a set of fixed text values to choose from. Right: In other cases, the CNL allows the user to enter certain values, such as the date.}
\label{fig:s4}
\end{figure}

\begin{figure}[h!]
\vspace*{-10mm}
\centering
\fbox{\includegraphics[width=\textwidth]{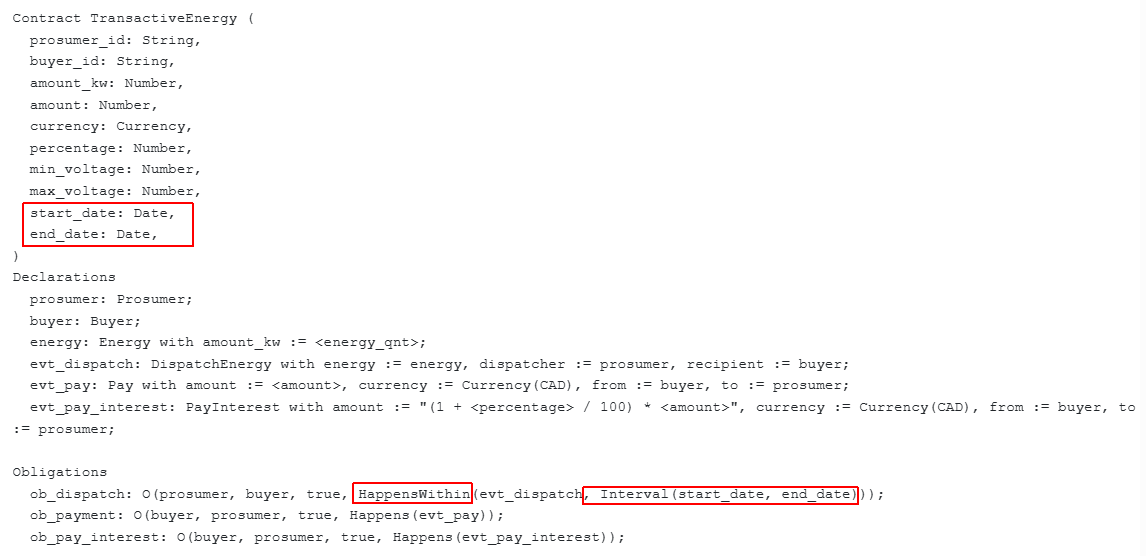}}
\caption{The \Symbo code is now updated with the required changes (to the domain model and event monitoring predicate) introduced by the refinement.}
\label{fig:s8}
\end{figure}

The first adjunct example is often found in a typical TE contract, so we will use this as an example. Traditional contract template programs may simply allow a user to fill in the two dates, but \SymbNLP allows a contract author a higher degree of customizability in how the contract is refined, by accommodating these more complex grammatical structures. Once the user enters and submits the refinement (see typical sequence in Figures~\ref{fig:s3} to~\ref{fig:s8}), \SymbNLP will automatically formalize the refined NL contract into the corresponding \Symbo specification. This specification is now ready to be used in the \SymbWeb project, as we illustrate in Section~\ref{FromSymbToSC}.

\vspace*{-8mm}
\section{\Symbo Editing and SC Code Generation}
\label{FromSymbToSC}

\begin{figure}[h!]
\centering
\fbox{\includegraphics[width=\textwidth]{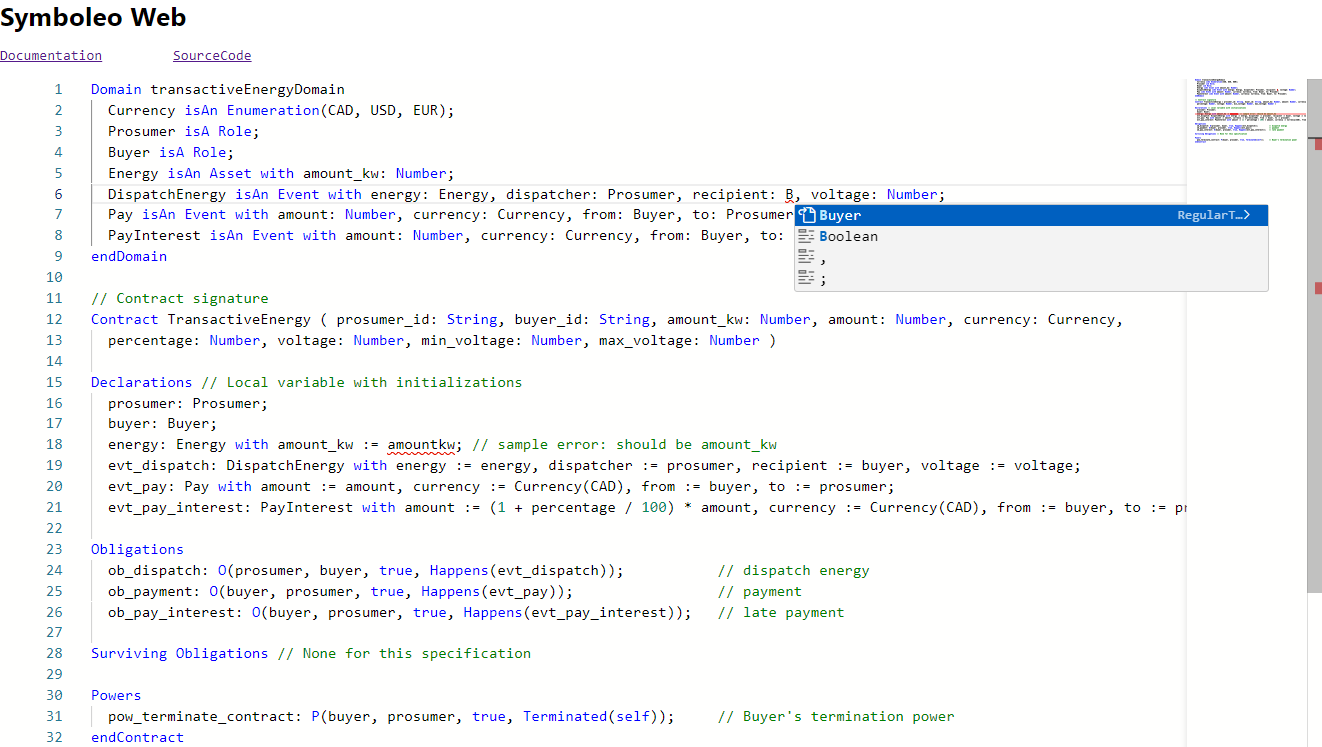}}
\caption{Online \SymbWeb editor, with syntax highlight, error reporting, and code completion for \Symbo specifications.}
\label{fig:SymboleoWeb}
\centering
\end{figure}

\begin{figure}[t]
\centering
\fbox{\includegraphics[width=\textwidth]{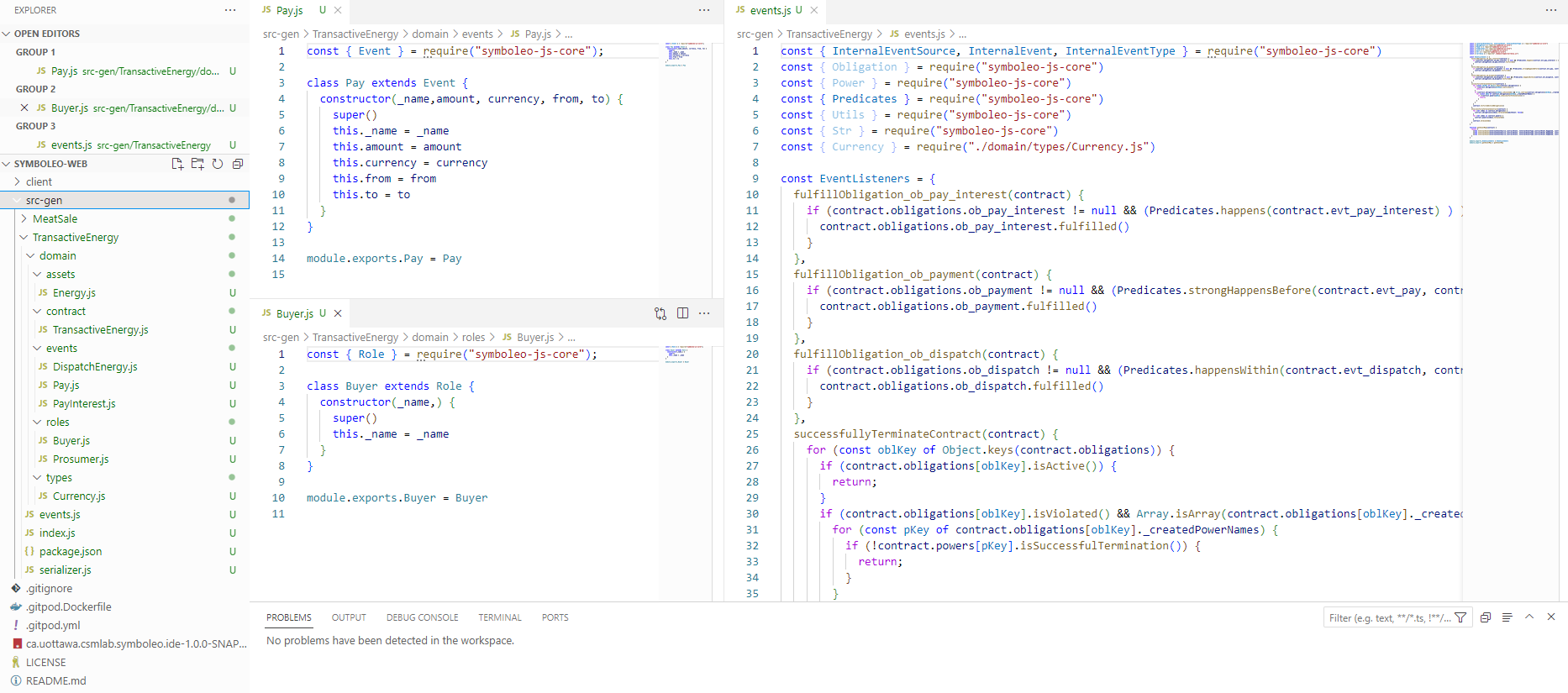}}
\caption{\SymbWeb editor showing the generated smart contract code (in JavaScript) for the initial \Symbo specification (Fig.~\ref{fig:SymboleoWeb}), with example files for an event, a role, and the event management.}
\label{fig:SmartContract}
\centering
\end{figure}

This section discusses \SymbWeb, our new web-based environment to help edit \Symbo specifications and generate smart contracts from them. Figures~\ref{fig:SymboleoWeb} and~\ref{fig:SmartContract} illustrate this tool. The code and a deployed version are available online\footnote{\url{https://bit.ly/Symboleo_Web}}. 

\SymbWeb primarily uses Microsoft's Monaco editor\footnote{\url{https://microsoft.github.io/monaco-editor/}}, which has become a cornerstone of various applications, most notably serving as the code editor for Visual Studio Code (VSCode). The editor supports a myriad of programming languages, boasts features such as syntax highlighting, autocompletion, and error diagnostics, and is renowned for its responsive user interface. This is combined with LSP4J \footnote{\url{https://github.com/eclipse-lsp4j/lsp4j}}, a Java implementation of the \textit{Language Server Protocol} (LSP), to be used by language servers built using Java. The LSP defines the communication protocol used between an editor/IDE and a language server. By complying with LSP, the server can be used by any client (VSCode desktop application, Web IDE, etc.). The \textit{WebSocket Secure} (WSS) protocol\footnote{\url{https://websockets.spec.whatwg.org/}} is also used to facilitate communication between the client and the language server. 

Figure~\ref{fig:SymboleoWeb} illustrates a TE \Symbo specification as seen in \SymbWeb. In this example, the contract specification contains an error in line 18, which is correctly highlighted as such. Furthermore, hovering over an error gives useful insights into how the error might be resolved. For example, in line 6, the editor suggests how to properly auto-complete a parameter type that starts with \texttt{\small B} (user-defined types are also suggested). The implementation of the validation and auto-completion rules defined by Rasti~\cite{aidin-thesis} contributes to improving the productivity of the user manually editing \Symbo specifications. 

Finally, as shown in Fig.~\ref{fig:SmartContract}, once the server recognizes an error-free \Symbo specification, it automatically invokes \SymbSC~\cite{rasti2024-Symboleo2SC} to generate the corresponding smart contract code in JavaScript, including the different assets, roles, obligations, events, etc. The user can view (and even edit, although this is not recommended) the generated files directly in \SymbWeb or (more likely) download them for direct deployment over the Hyperledger Fabric platform.


While \SymbWeb can be used to build a \Symbo contract from scratch, it can be especially useful when used in concert with \SymbNLP, as it exemplifies the flow from NL refinement to \Symbo specification, and then to executable code. \SymbNLP relies on various natural language processing and machine learning techniques, some of which are probabilistic~\cite{Meloche2023RE}, and can lead to slight errors in the \Symbo generation. \SymbWeb provides easy and intuitive means for a human user to correct these mistakes. 



\section{Analysis}
\label{Analysis}
We conducted an empirical analysis to assess the impact of NL template refinements on the resulting \Symbo specifications and smart contracts in JavaScript (JS). This also indirectly helps assess potential time and effort saved. We applied five distinct refinements, and four combinations thereof, to an initial NL contract template with two refinement locations ([P1] and [P2] in  Table~\ref{tab:contract}):
\begin{enumerate}
\item R1 $(P_{1}) = $ between [START\_DATE] and [END\_DATE]
\item R2 $(P_{2}) = $ before March 31, 2024 
\item R3 $(P_{2}) = $ within 2 weeks of Prosumer dispatching energy
\item R4 $(P_{1}) = $ before March 31, 2024
\item R5 $(P_{1}) = $ within 2 weeks of Buyer paying Prosumer.
\item R1R2 $(P_{1}, P_{2})$: \textit{Refinements R1 and R2} 
\item R1R3 $(P_{1}, P_{2})$: \textit{Refinements R1 and R3} 
\item R4R2 $(P_{1}, P_{2})$: \textit{Refinements R4 and R2}
\item R4R3 $(P_{1}, P_{2})$: \textit{Refinements R4 and R3}
\end{enumerate}

First, an initial \Symbo specification of the TE contract was created from a broadly specified NL contract and its corresponding specification (Sect.~\ref{FromNLtoSymboleo}). \SymbNLP was then used to generate our 9 refinements of the contract, each as a \Symbo file. These 9 specifications, together with the initial one, were fed to \SymbWeb to generate the corresponding 10 smart contracts. Table~\ref{tab:RefP} shows the results of our empirical analysis, where we measure the lines of code (LOC) that were added, modified, and deleted by several refinements to the initial \Symbo contract and corresponding JS smart contract. The LOC metric is often used as a proxy measure of development time and effort.

\newcommand*\rot{\rotatebox{90}}
\newcommand*\stack[2]{\vbox{\hbox{\strut \textbf{#1}}\hbox{\strut \textbf{#2}}}}

\begin{table}[b!]
    \centering
    \scriptsize
     \caption{Line of Code (LOC) counts and other statistics for the generated \Symbo specifications and smart contracts for five refinements (R1--R5) of parameters (P1 or P2) and four combinations thereof applied to the TE contract template from Table~\ref{tab:contract}.}
    \begin{tabular}{|l|*{10}{c|}}
        \hline
         \textbf{Refinements:} & {\textbf{Init}} & \stack{R1}{(P1)} & \stack{R2}{(P2} & \stack{R3}{(P2)} & \stack{R4}{(P1)} & \stack{R5}{(P1)} & \stack{R1R2}{(P1,P2)} & \stack{R1R3}{(P1,P2)} & \stack{R4R2}{(P1,P2)} & \stack{R4R3}{(P1,P2)}\\
        \hline
        \rowcolor{lightgreen}
        \multicolumn{11}{|l|}{\bf \Symbo}  \\   \hline
         \#LOC  & 40 & 40 & 40 & 41 & 40 & 42 & 40 & 41 & 40 & 41 \\
         \hline
        \#LOC added & & 0 & 0 & 1 & 0 & 2 & 0 & 1 & 0 & 1 \\
        \hline
     \#LOC modified & & 2 & 2 & 3 & 2 & 1 & 3 & 4 & 3 & 6 \\
        \hline
        \#LOC deleted & & 0 & 0 & 0 & 0 & 0 & 0 & 0 & 0 & 0 \\
        \hline
        \rowcolor{lightgreen}
        \multicolumn{11}{|l|}{\bf Smart Contract (SC), in JavaScript (JS), for Hyperledger Fabric}\\   
        \hline
        No. of file  & 12 & 12 & 12 & 12 & 12 & 13 & 12 & 12 & 12 & 12 \\
        \hline
        \#LOC SC & 606 & 608 & 607 & 630 & 607 & 647 & 609 & 632 & 608 & 631 \\
        \hline
        \#LOC added &  & 2 & 3 & 32 & 3 & 41 & 9 & 34 & 8 & 33 \\
        \hline
        \#LOC modified &  & 11 & 12 & 5 & 11 & 4 & 6 & 9 & 6 & 9 \\
        \hline
        \#LOC deleted & & 0 & 2 & 8 & 2 & 0 & 6 & 8 & 6 & 8 \\
        \hline
        \%LOC changed &  & 2.1\% & 2.5\% & 7.1\% & 2.6\% & 7.0\% & 3.4\% & 8.1\% & 3.3\% & 7.9\% \\
        \hline
        \rowcolor{lightgreen}
        \bf Ratio SC/Sym & 15.2 & 14.5 & 14.8 & 15.4 & 14.5 & 15.8 & 14.2 & 14.7 & 14.5 & 15.0 \\
        \hline
    \end{tabular}
    \label{tab:RefP}
\end{table}

As shown in Table~\ref{tab:RefP}, the \Symbo specifications of the contract and its refinements contain $\sim$41 LOC, while on average the corresponding smart contracts contain $\sim$618 LOC, resulting in an average LOC expansion ratio of about 1:15 (excluding the $\sim$3,000 LOC of JS code composing the \Symbo library, invoked by the generated code). The percentage of changed JS LOC varies from 2.1\% to 8.1\%, with the highest amount of changes corresponding to the more complex refinements (based on R3 and R5). These findings collectively provide insights into the nature of refinements in the context of \Symbo contracts. As these changes also impact multiple files, doing this manually would be both a time-consuming and error-prone maintenance activity. The sustained consistency in the ratios between specification sizes and generated code sizes also suggests relatively predictable time and effort savings over manual approaches. Finally, such savings grow with the number and complexity of NL refinements applied.

\section{Discussion}
\label{Discussion}
Methods and tools for helping in the smart contract creation, validation, and performance-monitoring have an immediate application to platforms where formalized contract templates with fixed variations are already in use, such as the mentioned transactive energy domain. In this area, the need is quite real, to support the practice of rapidly switching between power sources according to unpredictable swings in supply and demand. Performance could be automatic, including for example the payment of penalties from funds set aside for this purpose. Other concrete examples of applicable fields include:
\begin{itemize}
    \item \textbf{Online E-commerce platforms:} These platforms are now limited to fixed and extremely simple contractual formats, with very few choices. One can envisage marketplaces where one of the parties, usually the seller, would offer a wider range of choices, to be selected by the other party. For example, different `powers' can be offered to the other party to cover different circumstances such as the possibility of delivering certain products rather than others, or proposing different delivery options for different payments. The contract performance monitor would guide the parties through the steps necessary to realize the selected scenario.
    \item \textbf{Supply chain monitoring:} Throughout the supply chain, perishable food must adhere to specific standards, for purposes of health and traceability. Certain contracts may require that the food being transported stay within a specific temperature range, or that the food be delivered under specific conditions and/or within a specific time frame. Sensors and tracking devices could be configured to communicate with the smart contract, notifying it of events that are related to the fulfillment or violation of certain obligations.
\end{itemize}


Several approaches that formalize norms and enable their validation, including eFLINT~\cite{vanBinsbergen2022-eFlint,eFlint-2021} (which also offers a Web-based interface), only provide partial solutions to the above scenarios, whereas our Web-based environment for \Symbo goes further, with deployable monitoring smart contracts. 

At this point, \SymbNLP supports the formalization of a variety of temporal and conditional refinements, but future work may involve expanding the functionality to formalize more complex semantics. In most cases, the \Symbo code generated by \SymbNLP requires a small amount of manual revision and clean up, primarily minor formatting issues with the contract parameters. While the controlled NL used in \SymbNLP allows for the specification of simple events, there are some important limitations. The events that can be represented are restricted to simple events in an active form. Future work involves empirically deriving a more sophisticated event specification based on real contract data, which may include subordinate clauses, negations, and conjunctions. The refinements used in the  analysis were all temporal in nature, but we can also imagine the parameters being filled with other types of adjuncts, such as \textit{conditionals}. For example, we may want to specify that the energy must be delivered to the buyer \textit{if the market price for energy exceeds a predetermined threshold set by the buyer}. While these types of operations are supported by \SymbNLP, they are limited by the simplified specification of events. 

It may also be useful, particularly in the transactive energy domain, to specify state-based conditionals in the contract. For example, if the voltage of the dispatched energy is not within certain bounds, then there might be a financial penalty for the prosumer. While \SymbNLP includes conditional-based refinements, these are restricted to conditionals that include events. Future work may involve allowing for NL refinements that reflect \textit{state-based} conditions.

\SymbNLP currently operates with a CNL, since this allows for more predictable mappings to the formal specification. We may want to relax the refinements to allow for more freeform input, which would in turn allow for more expressiveness for the contract author. The trade-off for that expressiveness would be a reduction in its ability to properly formalize the refinement. However, future work may involve exploring the application of Large Language Models (LLMs) to this problem, which could potentially mitigate this reduction.

\section{Conclusion}
\label{Conclusion}
Legal contracts play a crucial role in business transactions between parties, and violating their obligations can lead to serious consequences. Smart contracts have been introduced to effectively monitor the performance providing early alerts for potential violations in contract performance. This paper presents practical and accessible solutions for formalizing contract templates and addressing challenges associated with template refinement and monitoring mechanisms. 

Our practical approach involves two complementary Web-based tools: \SymbNLP utilizes temporal and conditional refinements to convert NL templates into \Symbo specifications. Subsequently, these specifications are processed by the second tool, \SymbWeb, which validates and transforms them into smart contracts using \SymbSC. We evaluated the tools with a transactive energy contract example and an empirical analysis involving multiple refinements. Although our results are currently limited to one template, the key finding indicates a stable refinement process, as evidenced by a consistent ratio between specifications and generated code, and percentages of changes that grow with the number and complexity of NL refinements, which suggest corresponding time and effort savings as our approach is automated. 

In addition to the points raised in the previous section, future work involves a tighter integration of the two tools and of the \SymbPC property checker~\cite{Parvizimosaed2022-SymboleoPC}, additional performance experiments and improvements, further extensions of the smart contract code generator to better support security and other non-functional aspects, and formal usability studies involving real users.   



%
%
\bibliographystyle{splncs04}
\bibliography{SymboleoWebIDE}

\end{document}